\documentclass[final]{ws}

\newcommand{\tr}{\mathop{\rm tr}\nolimits}
\newcommand{\inst}{\mathop{\rm inst}\nolimits}
\newcommand{\T}{\mathop{\rm T}\nolimits}
\newcommand{\NP}{\mathop{\rm NP}\nolimits}
\newcommand{\PP}{\mathop{\rm P}\nolimits}
\newcommand{\eff}{\mathop{\rm eff}\nolimits}

\begin{document}

\title{Instanton infra-red stabilization in the nonperturbative QCD vacuum
\footnote{\uppercase{T}his work is supported by \uppercase{INTAS}
grant \uppercase{N}~110 and \uppercase{RFFI} grant 00-02-17836}}

\author{N.~O. Agasian and S.~M. Fedorov\footnote{\uppercase{W}ork partially
supported by \uppercase{INTAS} grant \uppercase{YS}~4433}}

\address{
Institute of Theoretical and Experimental Physics,\\
117218, Moscow, B.Cheremushkinskaya 25, Russia}

\maketitle

\abstracts{
The influence of nonperturbative fields on instantons
in quantum chromodynamics is studied. Nonperturbative vacuum is
described in terms of nonlocal gauge invariant vacuum averages of
gluon field strength. Effective action for instanton is derived in
bilocal approximation and it is demonstrated that stochastic
background gluon fields are responsible for infra-red (IR)
stabilization of instantons. Comparison of obtained instanton
size distribution with lattice data is made.}

\noindent {\bf 1.}
Instantons were introduced in 1975 by Polyakov and coauthors~\cite{BPST}.
Instanton gas as a model of QCD vacuum was proposed in works by
Callan, Dashen and Gross~\cite{CDG_76,CDG_78}. Instanton liquid model of
QCD vacuum was developed by Shuryak~\cite{Shuryak_81,Shuryak:nr}.
These topologically nontrivial field configurations are essential
for the solution of some problems of quantum chromodynamics.
Spontaneous chiral symmetry breaking (SCSB) can be explained with
the help of instanton and anti-instanton field configurations in
QCD vacuum~\cite{Diak_Pet_86}. Taking into account instantons is
of crucial importance for many phenomena of QCD
(see~\cite{Scha_Shur_98} and references therein).

At the same time, there is a number of problems in instanton physics.
The first is the divergence of integrals over instanton size $\rho$ at big $\rho$.
This makes it impossible to calculate instantons' contribution to some physical
quantities, such as vacuum gluon condensate. Second, ''area law'' for Wilson loop
can not be explained in instanton gas model, hence quasiclassical instanton
anti-instanton vacuum lacks confinement which is responsible for hadron spectra.

In our approach we make a natural assumption that there exist other nonperturbative
fields apart from instantons in the vacuum, which allow to explain listed above problems.
In this talk we will demonstrate that instanton can be stabilized in
nonperturbative vacuum and exist as a stable topologically nontrivial field
configuration against the background of stochastic nonperturbative fields, which
are responsible for confinement, and will find quantitatively it's size.

\noindent {\bf 2.}
Standard euclidian action of gluodynamics has
the form
\begin{equation}
\label{eq_action} S[A]=\frac{1}{2g_0^2} \int d^4 x
\tr(F_{\mu\nu}^2[A])= \frac{1}{4} \int d^4 x
F_{\mu\nu}^a[A]F_{\mu\nu}^a[A],
\end{equation}
where $ F_{\mu\nu}[A]=\partial_{\mu}A_{\nu} -
\partial_{\nu}A_{\mu}-i[A_{\mu},A_{\nu}]$ is the strength of gluon
field. We decompose $A_{\mu}$ as $A_{\mu} = A_{\mu}^{\inst}+B_{\mu}+a_{\mu}$,
where $A_{\mu}^{\inst}$ is an instanton-like field configuration
with a unit topological charge $Q_{\T}[A^{\inst}]=1$; $a_{\mu}$ is
quantum field and $B_{\mu}$ is nonperturbative background field (with zero
topological charge), which can be parametrized by gauge invariant
nonlocal vacuum averages of gluon field strength (correlators).

In general case effective action for instanton in NP vacuum takes
the form
\begin{equation}
\label{eq_genrl} Z=e^{-S_{\eff}[A^{\inst}]} = \int
[Da_{\mu}]\left\langle e^{-S[A^{\inst}+B+a]}\right\rangle,
\end{equation}
where $\langle...\rangle$ implies averaging over background field
$B_{\mu}$.

Integrating over $a_{\mu}$~\cite{Agas_Sim_95,Agasian_96} we arrive at
the following expression for effective instanton action:
\begin{equation}
\begin{array}{rl}
\label{eq_seff_gnrl}
&S_{\eff}[A^{\inst}]=S^{\PP}_{\eff}[A^{\inst}]+S^{\NP}_{\eff}[A^{\inst}]\\
&S^{\PP}_{\eff}(\rho)=\frac{b}{2}\ln\frac{1/\rho^2+m_*^2}{\Lambda^2}\\
&S^{\NP}_{\eff}[A^{\inst}]=-\ln\langle
Z_2(B)\rangle= -\ln\left\langle \exp
\{-S[A^{\inst}+B]+S[A^{\inst}]\}\right\rangle
\end{array}
\end{equation}

\begin{figure}[ht]
\centerline{\epsfxsize=2.3in\epsfbox{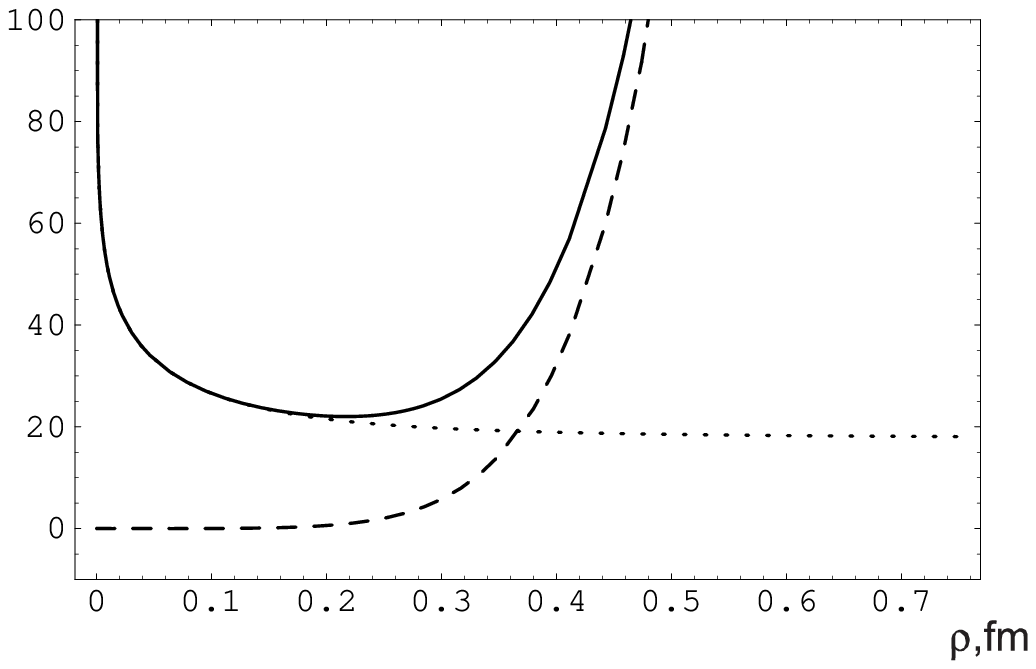}}
\caption{Effective action $S_{\eff}^{\PP}$ (dotted line),
$S^{\NP}_{\eff}$ (dashed line) and
$S_{\eff}=S_{\eff}^{\PP}+S^{\NP}_{\eff}$ (solid line) }
\label{fig3}
\end{figure}

\begin{figure}[ht]
\centerline{\epsfxsize=2.3in\epsfbox{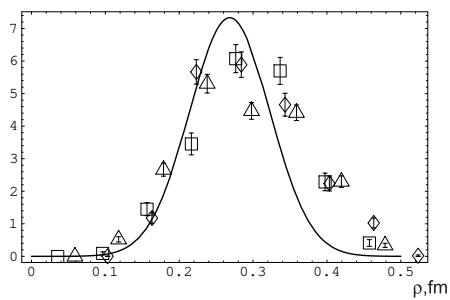}}
\caption{Instanton density $dn/d^4 z d\rho$ and lattice data~$^{12}$}
\label{fig4}
\end{figure}

\noindent{\bf 3.}
We calculated effective instanton action (see~\cite{AF_JHEP}) in bilocal approximation (i.e.
we considered only bilocal correlator). In many cases this approximation
appears to be sufficient for qualitative description of
various physical phenomena in QCD. Moreover, there are indications that corrections
due to higher correlators are small and amount to several percent~\cite{DShS}.
Numerical results for effective instanton action are
shown in Fig.~\ref{fig3}. It is clear that nonperturbative
part of effective action $S_{\eff}^{\NP}$ leads to IR stabilization
of instanton. Numerical results for instanton size distribution $dn/d^4 z d\rho \sim
\exp(-S_{\eff})$ and corresponding lattice
data~\cite{Hasen_Niet_98} are presented in Fig.~\ref{fig4}.
We can make a conclusion that our results for $\rho_c$ are consistent with phenomenological value
$\bar \rho \simeq 1/3$~fm~\cite{Shuryak_81,Shuryak:nr} and with lattice data.
We have also studied dependence of instanton size on gluon condensate $\langle G^2 \rangle$ and
correlation length $T_g$. We have found that, as it should be, big values of $\langle G^2 \rangle$
and $T_g$ lead to the suppression of large-size instantons.

\end{document}